\documentclass[letterpaper,conference]{IEEEtran}
\usepackage{ifthen}
\usepackage[cmex10]{amsmath} 

\usepackage{amsfonts}
\usepackage{amssymb}
\usepackage{lscape}
\usepackage{epsf}

\newtheorem{theorem}{\indent Theorem}[section]
\newtheorem{lemma}[theorem]{\indent Lemma}
\newtheorem{corollary}[theorem]{\indent Corollary}

\newtheorem{EXAMPLE}{\indent Example}[section]
\newtheorem{definition}{\indent Definition}[section]

\newenvironment{example}{\begin{EXAMPLE}\rm}{\rm\end{EXAMPLE}}

\newcommand{\code}{{\mathcal{C}}}
\newcommand{\graph}{{\mathcal{G}}}
\newcommand{\cL}{{\mathcal{L}}}
\newcommand{\cV}{{\mathcal{V}}}
\newcommand{\cH}{{\mathcal{H}}}
 
\newcommand{\cE}{{\mathcal{E}}}

\newcommand{\cM}{{\mathcal{M}}}

\newcommand{\cI}{{\mathcal{I}}}
\newcommand{\cJ}{{\mathcal{J}}}
\newcommand{\cN}{{\mathcal{N}}}
\newcommand{\cK}{{\mathcal{K}}}
\newcommand{\cP}{{\mathcal{P}}}

\newcommand\rr{{\mathbb R}}
\newcommand\zz{{\mathbb Z}}

\newcommand{\bldc}{{\mbox{\boldmath $c$}}}

\newcommand{\bldp}{{\mbox{\boldmath $p$}}}
\newcommand{\bldpp}{{\mbox{\scriptsize \boldmath $p$}}}

\newcommand{\bldv}{{\mbox{\boldmath $v$}}}

\newcommand{\bldx}{{\mbox{\boldmath $x$}}}
\newcommand{\bldy}{{\mbox{\boldmath $y$}}}
\newcommand{\bldyy}{{\mbox{\scriptsize \boldmath $y$}}}

\newcommand{\ones}{{\mbox{\boldmath $1$}}}
\newcommand{\zeros}{{\mbox{\boldmath $0$}}}

\newcommand{\rrr}{\mathfrak{R}}%
\newcommand{\rrrm}{\rrr^{-}}

\newcommand{\bldone}{{\mathbf{1}}}
\newcommand{\half}{{\textstyle\frac{1}{2}}}

    \def\squarebox#1{\hbox to #1{\hfill\vbox to #1{\vfill}}}

\newlength{\Algwidth}

\renewcommand{\baselinestretch}{1.0}

\begin{document}

\title{Lower Bounds on the Minimum Pseudodistance \\ for Linear~Codes \\ with~$q$-ary~PSK~Modulation~over~AWGN
}
\author{
\IEEEauthorblockN{Vitaly Skachek}
\IEEEauthorblockA{Claude Shannon Institute, University College Dublin \\
Belfield, Dublin 4, Ireland \\
Email:vitaly.skachek@ucd.ie} \\
\and
\IEEEauthorblockN{Mark F. Flanagan}
\IEEEauthorblockA{DEIS, University of Bologna \\
Via Venezia 52, 47023 Cesena, Italy \\
Email: mark.flanagan@ieee.org}
}

\maketitle

\begin{abstract}
We present lower bounds on the minimum pseudocodeword effective Euclidean distance (or minimum ``pseudodistance") for coded modulation systems using linear codes with $q$-ary phase-shift keying (PSK) modulation over the additive white Gaussian noise (AWGN) channel. These bounds apply to both binary and nonbinary coded modulation systems which use direct modulation mapping of coded symbols. The minimum pseudodistance may serve as a first-order measure of error-correcting performance for both linear-programming and message-passing based receivers. In the case of a linear-programming based receiver, the minimum pseudodistance may be used to form an exact bound on the codeword error rate of the system.
\vspace{2ex}

\end{abstract}

\section{Introduction}
\subsection{Background}

In classical coding theory, maximum-likelihood (ML) decoding of a signal-space code leads to a \emph{nearest-neighbour} decision rule in the signal space. For this reason, the \emph{minimum Euclidean distance} between modulated codewords (signal points) of a signal-space code is used as a first-order measure of its error-correcting performance under ML decoding. In the case of binary modulation, the minimum Hamming distance of the underlying code may be substituted, since in this case the Hamming distance is proportional to the squared Euclidean distance.

Recently, low-density parity check (LDPC) codes~\cite{Gallager} have attracted much interest due to 
their practical efficiency. In particular, it was shown that several families of LDPC codes can attain the capacity of various channels, when decoded by iterative \emph{message passing} (MP)
algorithms (for instance see~\cite{LMSS},~\cite{Shokrollahi},~\cite{Urbanke2}).  

The MP decoding algorithm operates locally on the \emph{Tanner graph}, a graph which represents the parity-check matrix. 
The notion of \emph{computation tree pseudocodewords} was introduced in~\cite{Wiberg} in order to adequately explain the limitations of MP decoding of binary LDPC codes. Computation tree pseudocodewords are closely related to \emph{graph-cover pseudocodewords}. The latter were extensively studied in~\cite{KV-characterization},~\cite{KV-IEEE-IT},~\cite{KV-Turbo} and~\cite{KV-lower-bounds}. The graph-cover pseudocodewords lie inside a region called the \emph{fundamental cone} (see~\cite{KV-characterization},~\cite{KV-IEEE-IT}). 
The set of graph-cover pseudocodewords were shown to be equivalent to the set of \emph{linear-programming (LP) pseudocodewords} for the cases of binary~\cite{Feldman-thesis},~\cite{Feldman} and nonbinary coded modulation systems~\cite{FSBG},~\cite{SFBG-2}. In both binary and nonbinary cases, necessary and sufficient conditions for codeword error under linear programming (LP) decoding could be expressed is terms of these LP pseudocodewords, assuming transmission of the all-zero codeword (\cite{Feldman},~\cite{FSBG}). 

In~\cite{FKKR}, the pseudocodeword effective Euclidean distance, or \emph{pseudodistance}, is associated with any pseudocodeword. This concept of pseudodistance was shown in~\cite{FKKR} to play an analagous role to that of the signal Euclidean distance in ML decoding. The minimum pseudodistance is defined as the minimum over all pseudodistances of pseudocodewords; this may be taken as a first-order measure of decoder error-correcting performance for LP or MP decoding. In~\cite{KV-lower-bounds}, it was shown that bounds could be obtained on the minimum pseudoweight of a binary linear code, these bounds being expressed in terms of parameters of the parity-check matrix of the code.

In this work, we extend the results in~\cite{KV-lower-bounds} to the nonbinary case. In particular, we
show that bounds on the minimum pseudodistance can be obtained for the case of nonbinary coding and modulation, which are generalizations of the bounds on pseudoweight in~\cite{KV-lower-bounds} for the case of binary coding and modulation. 
Generally, the techniques are based on the techniques therein, although some additional ideas are used.  

\subsection{Basic Definitions}
\label{sec:defs}

We consider codes over finite rings (this includes codes over finite fields, but may be more general). 
Denote by $\rrr$ a ring with $q$ elements, by $0$ its additive identity, and let $\rrrm = \rrr \backslash \{ 0 \}$. 
Let $\code$ be a linear $[n,k]$ code with parity-check matrix $\cH$ over~$\rrr$
(we assume that $\rrr$ is quasi-Frobenius, which implies that the parity-check matrix exists). 
The parity check matrix $\cH$ has $m \ge n-k$ rows.

Denote the set of column indices and the set of row indices of $\cH$ by  $\cI = \{1, 2, \cdots, n \}$ 
and $\cJ = \{1, 2, \cdots, m \}$, respectively. 
We use notation $\cH_j$ for the $j$-th row of $\cH$, and $\cI_j$ for the support of $\cH_j$.  


Given any $\bldc \in \rrr^n$, we say that parity check $j\in\cJ$ is \emph{satisfied} by $\bldc$ if and only if
\begin{equation}
\sum_{i\in\cI_j} c_i \cdot \cH_{j,i} = 0 \; .
\label{eq:parity_check_satisfied}
\end{equation}
Also, we say that the vector $\bldc$ is a codeword of $\code$, writing $\bldc \in \code$, if and only if all parity checks $j\in\cJ$ are satisfied by $\bldc$.

Let the graph $\graph = (\cV, \cE)$ be the Tanner graph of $\code$ associated with the parity-check matrix $\cH$. This graph has vertex set $\cV = \{u_1, u_2, \cdots, u_n \} \cup  
\{v_1, v_2, \cdots, v_m \}$, and there is an edge between $u_i$ and $v_j$ if and only if $\cH_{j,i} \neq 0$. This edge is labelled with the value $\cH_{j,i}$. 
We denote by $\cN(v)$ the set of neighbors of a vertex $v\in\cV$. For a word $\bldc  = (c_1, c_2, \cdots, c_n) \in \rrr^n$, 
we associate the value $c_i$ with variable vertex $u_i$ for each $i \in \cI$. 
It may be easily seen that the Tanner graph provides a graphical means of checking whether each parity-check $j\in\cJ$ is satisfied, and hence whether the vector $\bldc$ is a codeword of $\code$. 

\begin{definition}
(\cite{KV-characterization})
A graph $\tilde{\graph} = (\tilde{\cV}, \tilde{\cE})$ is a \emph{finite cover} of the graph $\graph = (\cV, \cE)$ 
if there exists a mapping $\Pi: \tilde{\cV} \rightarrow \cV$ which is a graph homomorphism
($\Pi$ takes adjacent vertices of $\tilde{\graph}$ to adjacent vertices of $\graph$), such that 
for every vertex $v \in \graph$ and every $\tilde{v} \in \Pi^{-1}(v)$, the neighborhood $\cN(\tilde{v})$ 
of $\tilde{v}$ (including edge labels) is mapped bijectively to $\cN(v)$. 
\end{definition}

\begin{definition}
(\cite{KV-characterization})
A cover of the graph $\graph$ is called an $M$-cover, where $M$ is a positive integer, if $|\Pi^{-1}(v)| = M$
for every vertex $v \in \cV$.  
\end{definition}
 
Fix some positive integer $M$. Let $\tilde{\graph} = (\tilde{\cV}, \tilde{\cE})$ be an $M$-cover   
of the Tanner graph $\graph = (\cV, \cE)$ of the code $\code$ associated with the parity-check matrix $\cH$. 
Denote the vertices in the sets $\Pi^{-1} (u_i)$ and $\Pi^{-1} (v_j)$ by 
$\{ u_{i,1}, u_{i,2}, \cdots, u_{i,M} \}$ and $\{ v_{j,1}, v_{j,2}, \cdots, v_{j,M} \}$, respectively, where $i\in\cI$ and $j\in\cJ$.

Consider the linear code $\tilde{\code}$ of length $Mn$ over $\rrr$, 
defined by the $Mm \times Mn$ parity-check matrix $\tilde{\cH}$. For 
$1 \le i^*,j^* \le M$ and $i \in \cI$, $j \in \cJ$, we let  
$i' = (i-1) M + i^*, j' = (j-1) M + j^*$, and
\begin{equation}
\tilde{\cH}_{j',i'} = \left\{ \begin{array}{cl}
\cH_{j,i} & \mbox{if } u_{i,i^*} \in \cN(v_{j,j^*}) \\
0 & \mbox{otherwise} 
\end{array} \right. \; .
\label{eq:H-tilde} 
\end{equation}
Then, any vector $\bldp \in \tilde{\code}$ has the form 
\begin{eqnarray*}
 \bldp & = & ( p_{1,1}, p_{1,2}, \cdots, p_{1,M}, p_{2,1}, p_{2,2}, \\
 && \hspace{3ex} \cdots, p_{2,M}, \cdots, p_{n,1}, p_{n,2}, \cdots, p_{n, M} ) \; .
\end{eqnarray*}
We associate the value 
$p_{i,\ell} \in \rrr$ with the vertex $u_{i,\ell}$ in $\tilde{\graph}$ ($i\in\cI$, $\ell = 1, 2, \cdots, M$). It may be seen that $\tilde{\graph}$ is the Tanner graph of the code $\tilde{\code}$ associated with the parity-check matrix $\tilde{\cH}$.

The word $\bldp \in \tilde{\code}$ as above is called a \emph{graph-cover pseudocodeword} 
of the code $\code$. We also define the $n \times q$ \emph{pseudocodeword matrix} corresponding to $\bldp$ by 
\[
\cP = \Big( m_i^{(\alpha)} \Big)_{i \in \cI; \, \alpha\in\rrr} \; ,  
\] 
where 
\[
m_i^{(\alpha)} = \left| \{ \ell \in \{ 1, 2, \cdots, M \} \; : \; p_{i,\ell} = \alpha \} \right| \ge 0 \; , 
\]
for $i\in\cI$, $\alpha \in \rrr$. 
We then define the \emph{normalized pseudocodeword matrix} corresponding to $\bldp$ by 
\[
\cP_0 = \Big( f_i^{(\alpha)} \Big)_{i \in \cI; \, \alpha \in \rrr} \; , 
\] 
where $f_i^{(\alpha)} = m_i^{(\alpha)}/M$ for every $i \in \cI$, 
$\alpha \in \rrr$. 

In~\cite{FSBG,FSBG-2}, another set of pseudocodewords, called \emph{linear-programming pseudocodewords}, was defined. These LP pseudocodewords, which also admit a matrix representation, were shown to be directly linked to codeword error events in LP decoding. It was also shown in~\cite{FSBG,FSBG-2} that the two pseudocodeword concepts are equivalent, i.e. there exists an LP pseudocodeword with a particular pseudocodeword matrix if and only if there exists a graph-cover pseudocodeword having the same pseudocodeword matrix. 

It was shown in~\cite{FSBG} and \cite{Flanagan_cw_ind} that for the case of $q$-ary PSK transmission over AWGN under LP or MP decoding, 
the codeword error rate performance is independent of the transmitted codeword under the following conditions. First,  
$\rrr$ under addition is a cyclic group. If we let $\beta$ be a generator in $\rrr$ then we may write $\rrr = \{ 0, \beta, 2 \beta, \cdots, (q-1) \beta \}$ where $k \beta = \beta + \cdots + \beta$ ($k > 0$ terms in sum). Second, the modulation mapping is the `natural' mapping
\begin{equation}
\cM (k \beta) = \exp \left( \frac{\imath \cdot 2 \pi k}{q} \right) \; , 
\label{eq:modulation_mapping}
\end{equation}
where $\imath = \sqrt{-1}$. 
We assume in this work that these conditions hold; hence in the sequel, we adopt the simpler notation $f_i (k)$ for $f_i^{(k \beta)}$, $k=0,1,\cdots, q-1$. 
 
\section{Bounds on the Pseudodistance of Individual Pseudocodewords} 

In~\cite{Feldman}, for the case of binary coding and binary modulation, the set of pseudocodewords was used to characterize the error correction capability of the system under LP decoding. This was extended to the case of nonbinary coding and modulation in~\cite{FSBG}. In~\cite{FKKR}, it was observed that with each pseudocodeword $\bldp$ may be associated a point in the signal space; these signal points then play a role in LP decoding analagous to that of the modulated codewords in ML decoding. In particular, we may associate with each pseudocodeword an effective Euclidean distance from the modulated all-zero codeword, or pseudodistance, $d_{\textrm{eff}}(\bldp)$ (pseudodistance with respect to the all-zero codeword is sufficient assuming the symmetry condition above). Then, assuming LP decoding, the event $E(\bldp)$ = ``on transmission of the all-zero codeword, there is a codeword error due to pseudocodeword $\bldp$'' has probability
\begin{equation}
P(E(\bldp)) = Q\left(\frac{d_{\textrm{eff}}(\bldp)}{2 \sigma}\right)
\label{eq:Prob_of_CW_error_due_to_PCW}
\end{equation}
(where $\sigma^2$ is the noise variance per dimension, and $Q(x) = \frac{1}{2 \pi} \int_x^\infty \exp(-t^2/2) dt$ denotes the Gaussian $Q$-function) and thus the probability of codeword error is equal to $P(\bigcup E(\bldp))$ where the union is over the set of all pseudocodewords $\bldp$ (equation (\ref{eq:Prob_of_CW_error_due_to_PCW}) was stated in~\cite{FKKR} for the case of MP decoding and computation tree pseudocodewords). Therefore the minimum pseudodistance $d_{\textrm{min}} = \min_\bldpp \{ d_{\textrm{eff}}(\bldp) \}$ may be taken as a first-order measure of error-correcting performance of the coded modulation system. For the case of MP decoding and graph-cover pseudocodewords, (\ref{eq:Prob_of_CW_error_due_to_PCW}) may be taken as an approximation. Also, for the case of binary coding and modulation, the pseudocodeword effective Hamming weight (or ``pseudoweight") may be defined by $w_{\textrm{eff}}(\bldp) = d^2_{\textrm{eff}}(\bldp)/4$ by analogy with the case of classical ML decoding~\cite{FKKR}. 

It was shown in \cite{Kelley-Sridhara-ISIT-2006} that for the case of $q$-ary PSK modulation over AWGN, the squared pseudodistance between the all-zero codeword and the pseudocodeword $\bldp$ is given by
\begin{equation}
d^2_{q}(\bldp) = \frac{S^2}{V} \; , 
\label{eq:w-q-awgn}
\end{equation}
where 
\begin{equation}
S = 2 \sum_{i \in \cI} \left( 1-  \sum_{k=0}^{q-1} f_i(k) \cdot \cos \left( \frac{2 \pi k}{q} \right) \right) \; , 
\label{eq:M}
\end{equation}
and 
\begin{eqnarray}
V & = & \sum_{i \in \cI} \Bigg( \sum_{k=0}^{q-1} f^2_i(k) \nonumber \\
&& + \; 2 \; \mathop{\sum_{k < \ell}}_{ k, \ell \in \{0, \dots, q-1 \}} f_i(k) f_i(\ell) \cdot 
\cos \left( \frac{2 \pi (k - \ell)}{q} \right) \nonumber \\
&& - \; 2 \; \sum_{k=0}^{q-1} f_i(k) \cdot \cos \left( \frac{2 \pi k}{q} \right) + 1 \Bigg) \; . 
\label{eq:V}
\end{eqnarray}
By rearrangement of the expression in~(\ref{eq:M}), we have 
\begin{eqnarray}
S & = & 2 \sum_{i \in \cI} \left( 1- \sum_{k=0}^{q-1} f_i(k) \cdot \cos \left( \frac{2 \pi k}{q} \right) \right) 
\nonumber \\
& = & 2 \sum_{i \in \cI} \left( \sum_{k=1}^{q-1} f_i(k) \cdot \left( 1 - \cos \left( \frac{2 \pi k}{q} \right) \right) 
\right) \nonumber \\
& \ge & 2 \left( 1 - \cos \left( \frac{2 \pi}{q} \right) \right) \cdot \sum_{i \in \cI} \left( \sum_{k=1}^{q-1} f_i(k) \right)  
\; .  
\label{eq:M-q}
\end{eqnarray}
Similarly, for~(\ref{eq:V}) we have 
\begin{eqnarray}
V & = & \sum_{i \in \cI} \Bigg( \sum_{k=1}^{q-1} f^2_i(k) + f^2_i(0) \nonumber \\
&& + \; 2 \; \mathop{\sum_{k < \ell}}_{k, \ell \in \{1, \dots, q-1 \}} f_i(k) f_i(\ell) \cdot 
\cos \left( \frac{2 \pi (k - \ell)}{q} \right) \nonumber \\
&& \; + \; 2 \sum_{\ell = 1}^{q-1} f_i(0) f_i(\ell) \cdot 
\cos \left( \frac{2 \pi \ell}{q} \right) \nonumber \\
&& - 2 \sum_{k=1}^{q-1} f_i(k) \cdot \cos \left( \frac{2 \pi k}{q} \right) - 2 f_i(0) + 1 \Bigg) \nonumber \; . 
\end{eqnarray}
It follows that 
\begin{eqnarray}
V & \le &  \sum_{i \in \cI} \Bigg( \left(\sum_{k=1}^{q-1} f_i(k) \right)^2 \; + \; (1 - f_i(0))^2 \nonumber \\
&& \; + \; 2 \sum_{k = 1}^{q-1} f_i(k)  
\cos \left( \frac{2 \pi k}{q} \right) \cdot ( f_i(0) - 1 ) 
\Bigg) \; . \nonumber
\end{eqnarray}
After rearrangement, we obtain 
\begin{eqnarray}
V & \le & 4 \; \sum_{i \in \cI} \left(\sum_{k=1}^{q-1} f_i(k) \right)^2  \; . 
\label{eq:V-q}
\end{eqnarray}
We substitute the expressions in~(\ref{eq:M-q}) and~(\ref{eq:V-q}) into~(\ref{eq:w-q-awgn}), and obtain that 
\begin{equation}
d^2_{q}(\bldp) \ge \left( 1 - \cos ( 2 \pi / q) \right)^2 \cdot 
\frac{ \left( \sum_{i \in \cI} x_i \right)^2 } {\sum_{i \in \cI} x_i^2 } \; ,
\label{eq:w-q}
\end{equation}
where 
\[
  \forall i \in \cI \; : \; x_i =  \sum_{k=1}^{q-1} f_i(k)
      \; . 
\]

\begin{example}
Take $\rrr = \{ 0, 1 \}$ with binary signaling over AWGN. In this case, $q =2$, and 
(\ref{eq:w-q}) can be re-written as
\[
d^2_{2}(\bldp) \ge 4 \; \frac{ \left( \sum_{i \in \cI} x_i \right)^2 } {\sum_{i \in \cI} x_i^2 } \; , 
\]
which accords with the well-known pseudoweight expression for the case of binary code and modulation~\cite{KV-characterization}. 
\end{example}

\begin{example}
\label{ex:3-1}
Take $\rrr = \zz_3 =\{0, 1, 2\}$ with ternary PSK modulation over AWGN. We show that in this case
the inequality~(\ref{eq:w-q}) can be slightly improved. 
Observe that in this case,~(\ref{eq:M}) and~(\ref{eq:V}) can be re-written as
\begin{eqnarray}
S & = & 2 \sum_{i \in \cI} \left( \half f_i(1) + \half f_i(2) + (1 - f_i(0)) \right) \nonumber \\
& = & 3 \sum_{i \in \cI} \left( f_i(1) + f_i(2)\right) \; ,  
\label{eq:M-3}
\end{eqnarray}
and 
\begin{eqnarray}
V & = & 
\sum_{i \in \cI} \Big( 3 (f_i(1))^2 + 3 (f_i(2))^2 + 3 f_i(1) f_i(2) \Big) \nonumber \\
& \le & 3 \; \sum_{i \in \cI} \Big( (f_i(1)) + (f_i(2)) \Big)^2 \; , 
\label{eq:V-3}
\end{eqnarray}
where the last equality in~(\ref{eq:M-3}) and the equality in~(\ref{eq:V-3})
are due to the fact that $\sum_{k=0}^{q-1} f_i(k) = 1$ for all $i \in \cI$. 

Finally, we substitute the expressions in~(\ref{eq:M-3}) and~(\ref{eq:V-3}) into~(\ref{eq:w-q-awgn}) to obtain that 
\begin{equation}
d^2_{3}(\bldp) \ge 3 \cdot \frac{\left( \sum_{i \in \cI} x_i \right)^2}
{\sum_{i \in \cI} x_i^2} \; . 
\label{eq:w-3-awgn}
\end{equation}
\end{example}

\begin{example}
\label{ex:4-1}
Take $\rrr = \zz_4 =\{0, 1, 2, 3\}$ with quaternary PSK (QPSK) modulation over AWGN channel. 
In this case, using~(\ref{eq:w-q}), we obtain that 
\begin{equation}
d^2_{4}(\bldp) \ge \frac{ \left( \sum_{i \in \cI} x_i \right)^2 } 
{\sum_{i \in \cI} x_i^2 } \; ,
\label{eq:w-4-awgn}
\end{equation}
where 
\[
\bldx = (x_i)_{i=1}^n = ( f_i(1) + f_i(2) + f_i(3))_{i=1}^n \; . 
\]
\end{example}

\section{Inequalities for Pseudocodewords}

A complete characterization of the \emph{fundamental cone}, in which the pseudocodewords lie, was given for the case of binary coding and modulation in~\cite{KV-characterization}. For the present more general framework, a complete characterization of the corresponding fundamental region appears to be a difficult task. In this section we derive a set of inequalities which must be satisfied by the entries of any pseudocodeword matrix; these inequalities must necessarily be satisfied by any pseudocodeword lying in the fundamental cone. These inequalities will be helpful in deriving the bounds on minimum pseudodistance in the sequel. 

\begin{theorem}
Let $\code$ be a linear $[n,k]$ code over $\rrr$ with an $m \times n$ parity-check matrix $\cH$. 
Let $\cI$, $\cJ$ and $\cI_j$ be defined as in Section~\ref{sec:defs}. Assume
that $\cH_{j,i}$ is not a zero-divisor in $\rrr$ for any $j \in \cJ$, $i \in \cI_j$. 
Let 
\[
\cP = \left( m_i^{(\alpha)} \right)_{i \in \cI ; \alpha \in \rrr}  
\]
be the pseudocodeword matrix of a graph-cover pseudocodeword $\bldp$ of the code $\code$ with parity-check matrix $\cH$. Then, for any $j \in \cJ$, $\ell \in \cI_j)$, 
\begin{equation}
\sum_{i \in \cI_j \backslash \{ \ell \}} \;\; \sum_{b \in \rrrm} 
m_i^{(b)} \; \ge \; \sum_{b \in \rrrm } m_\ell^{(b)} \; .
\label{eq:thrm-ineq}
\end{equation}
\label{thrm:simple-inequality} 
\end{theorem}

\begin{proof} Suppose the graph-cover pseudocodeword $\bldp$ corresponds to the $M$-cover $\tilde{\graph} = (\tilde{\cV}, \tilde{\cE})$, and let $\tilde{\code}$ be the linear code of length $Mn$ over $\rrr$ defined by the parity-check matrix $\tilde{\cH}$ described by~(\ref{eq:H-tilde}). Then, $\tilde{\graph}$ is the Tanner graph of the code $\tilde{\code}$ associated with the parity-check matrix $\tilde{\cH}$.

Take some $j \in \cJ$ and $\ell \in \cI_j$. Fix some $1 \le j^* \le M$, and take the $j^*$-th 
copy $v_{j,j^*} \in \tilde{\cV}$ of the parity-check vertex $v_j \in \cV$. Let
\[
\left\{ u_{i,\sigma(i,j^*)} \right\}_{i \in \cI_j} = \cN(v_{j,j*}) \subseteq \tilde{\cV} \; ,
\]
where $1 \le \sigma(i,j^*) \le M$ for every $i \in \cI_j$. 

Denote $j' = (j-1)M + j^*$. Since $\bldp \in \tilde{\code}$, 
\[
\sum_{i \in \cI_j } \cH_{j', (i-1)M + \sigma(i,j^*)} \cdot p_{i, \sigma(i,j^*)} \; = \; 0 \; . 
\]
This can be rewritten as 
\begin{equation}
\sum_{i \in \cI_j} \cH_{j, i} \cdot p_{i, \sigma(i,j^*)} \; = \; 0 \; . 
\label{eq:thrm-parity}
\end{equation}
Assume that $p_{\ell, \sigma(\ell,j^*)} \neq 0$. Then, 
\begin{equation}
\sum_{i \in \cI_j  \backslash \{ \ell \}} \cH_{j, i} \cdot p_{i, \sigma(i,j^*)} 
\; = \; - \cH_{j, \ell} \cdot p_{\ell, \sigma(\ell,j^*)} \; ,
\label{eq:non-zero}
\end{equation}
and, since $\cH_{j, \ell}$ is not a zero divisor in $\rrr$, the expression in~(\ref{eq:non-zero}) is non-zero. 
Therefore, there exists at least one $i_{j^*} \in \cI_j$, $i_{j^*} \neq \ell$, such that
\[
p_{i_{j^*}, \sigma(i_{j^*},j^*)} \neq 0 \; . 
\] 

The number of indices $j^*$ ($1 \le j^* \le M$) such that $p_{\ell, \sigma(\ell,j^*)} \neq 0$ is given
by $\sum_{b \in \rrrm} m_\ell^{(b)}$. This number is equal to the number of indices $j^*$ ($1 \le j^* \le M$) such that 
$p_{i_{j^*}, \sigma(i_{j^*},j^*)} \neq 0$, which, in turn, is less than or equal to 
\[
\sum_{i \in \cI_j \backslash \{ \ell \}} \;\; \sum_{b \in \rrrm} 
m_i^{(b)} \; . 
\]
\end{proof}

On division of both sides of~(\ref{eq:thrm-ineq}) by $M$, we obtain the following result. 

\begin{corollary}
Let $\code$, $\cH$ and $\cP$ be defined as in Theorem~\ref{thrm:simple-inequality}. 
Then, for any $j \in \cJ$, $\ell \in \cI_j$, 
\begin{equation}
\sum_{i \in \cI_j \backslash \{ \ell \}} \;\; \sum_{k=1}^{q-1}
f_i(k) \; \ge \; \sum_{k=1}^{q-1} f_\ell(k) \; .
\end{equation}
\label{crlr:simple-inequality} 
\end{corollary}


\section{Eigenvalue Bound}
\label{sec:lower-bound-1}

In this section, we consider $(c,d)$-regular codes, i.e. the parity check matrix $\cH$ of $\code$ has $c$ nonzero elements per column and $d$ nonzero elements per row. Throughout this section, 
let $\code$ be a $(c,d)$-regular linear $[n,k]$ code over $\rrr$ with an $m \times n$ parity-check matrix $\cH$,
and assume that $\cH_{j,i}$ is not a zero-divisor in $\rrr$ for any $j \in \cJ$, $i \in \cI_j$. 
Let 
\[
\cP_0 = \left( f_i(k) \right)_{i \in \cI ; k \in \{ 1,2,\cdots, q-1 \} }  
\]
be the normalized pseudocodeword matrix of a graph-cover pseudocodeword $\bldp$ corresponding to some $M$-cover of 
the Tanner graph of $\cH$. 
Denote
\[
\forall i \in \cI \; : \; x_i =  \sum_{k=1}^{q-1} f_i(k) \qquad \mbox{and} \qquad  \bldx = (x_i)_{i \in \cI} \; . 
\]

We define a real matrix $\cL = \cH_s^T \cdot \cH_s$, where $\cH_s$ is an $m \times n$ matrix whose entries are equal to one on the support of $\cH$, and are equal to zero otherwise. We assume that the Tanner graph of $\code$ corresponding to $\cH$ consists of a single connected component, and denote by 
$\lambda_1 \ge \lambda_2 \ge \cdots \ge \lambda_n$ the eigenvalues of $\cL$. 
Let $\bldv_1, \bldv_2, \cdots, \bldv_n$ be the set of orthonormal eigenvectors corresponding to eigenvalues 
$\lambda_1, \lambda_2, \cdots, \lambda_n$ of the matrix $\cL$, respectively. 
Then, $\lambda_1 = c \cdot d > \lambda_2$, and $\bldv_1 = \frac{1}{\sqrt{n}} \cdot \bldone$. Also, assume that $q$-ary PSK modulation is used for transmission over the AWGN channel. 

Our analysis follows the lines of~\cite{KV-lower-bounds}.    
\begin{lemma}
Let $\cP_0$ and $\bldx$ be defined as above. 
Then, for any $j \in \cJ$, 
we have
\[
\left( \sum_{i \in \cI_j} \quad x_i \right) ^2 
\ge 2 \cdot \sum_{i \in \cI_j} x_i^2 \; .  
\]
\label{lemma:ineq-1-q}
\end{lemma}
\begin{proof} 
For any $j \in \cJ$ write
\begin{eqnarray*}
\left( \sum_{i \in \cI_j} x_i \right)^2 & = & 
\left( \sum_{i \in \cI_j} x_i \right)
\cdot \left( \sum_{\ell \in \cI_j} x_\ell \right) \\
& = & 
\sum_{i \in \cI_j} x_i 
\left( \sum_{\ell \in \cI_j} x_\ell \right) \\
& \ge & \sum_{i \in \cI_j} 2 \cdot x_i^2 \; , 
\end{eqnarray*}
where the inequality is due to Corollary \ref{crlr:simple-inequality}.  
\end{proof}

The following lemma is the counterpart of Lemma~8 in~\cite{KV-lower-bounds}. 
\begin{lemma}
Let $\bldx$ be a vector defined as in Lemma~\ref{lemma:ineq-1-q}, and let $\bldy = \bldx \cdot \cH_s^T$. Then, 
\[
|| \bldy ||_2^2 \; \ge \; 2c \cdot || \bldx ||_2^2 \; . 
\]
\label{lemma:ineq-2-q}
\end{lemma}
\vspace{-2ex}
\begin{proof}
\vspace{-2ex}
We write 
\[
|| \bldy ||_2^2 \; = \; \sum_{j \in \cJ} y_j^2 
\; = \; \sum_{j \in \cJ} \left( \sum_{i \in \cI_j} x_i \right)^2 \; .
\]
We apply Lemma~\ref{lemma:ineq-1-q} to obtain that  
\[
|| \bldy ||_2^2 \; \ge \; \sum_{j \in \cJ} \; 2 \cdot \sum_{i \in \cI_j} x_i ^2
\; = \; 2c \cdot || \bldx ||_2^2 \; ,
\]
where the last transition is due to the fact that each column in $\cH$ contains exactly $c$ nonzero symbols.
\end{proof}

The following lemma is based on Theorem~10 in~\cite{KV-lower-bounds}. 

\begin{lemma}
Let $\bldx$ and $\bldy$ be vectors defined 
as in Lemma~\ref{lemma:ineq-2-q}, and let $\lambda_1$ and $\lambda_2$ be defined as in Section~\ref{sec:lower-bound-1}.
Then, 
\[
|| \bldy ||_2^2 \le \frac{\lambda_1 - \lambda_2}{n} \left( \sum_{i \in \cI} x_i \right)^2 + \lambda_2 || \bldx ||_2^2 \; . 
\]
\label{lemma:ineq-q}
\end{lemma}

\begin{proof}
Write $\bldx$ as 
\[
\bldx = \sum_{i=1}^n \sigma_i \bldv_i \; , 
\]
where $\bldv_i$ ($i \in \cI$) are defined in Section~\ref{sec:lower-bound-1}, and 
$\sigma_i$ ($i \in \cI$) are real numbers. 
In particular, 
\[
\sigma_1 = \frac{1}{\sqrt{n}} \langle \bldx, \bldone \rangle = \frac{1}{\sqrt{n}} \left( \sum_{i=1}^n x_i \right) \; . 
\]

Then, 
\begin{eqnarray*}
|| \bldy ||_2^2 & = & || \bldx \cdot \cH_s^T ||_2^2 = \bldx \cdot \cL \cdot \bldx^T \\
& = & \sum_{i=1}^n \sigma_i \bldv_i \cdot \cL \cdot \sum_{i'=1}^n \sigma_{i'} \bldv_{i'} \\
& = & \sum_{i=1}^n \sigma_i \bldv_i \cdot \sum_{i'=1}^n \lambda_{i'} \sigma_{i'} \bldv_{i'} \\
& = & \sum_{i=1}^n \lambda_i \sigma_i^2 
\; = \; \frac{\lambda_1}{n} \left( \sum_{i=1}^n x_i \right)^2 + \sum_{i=2}^n \lambda_i \sigma_i^2 \\
\end{eqnarray*}
\begin{eqnarray*}
\phantom{|| \bldy ||_2^2} & \le & \frac{\lambda_1}{n} \left( \sum_{i=1}^n x_i \right)^2 + \lambda_2 \sum_{i=2}^n \sigma_i^2 \\
& = & \frac{\lambda_1}{n} \left( \sum_{i=1}^n x_i \right)^2 + \lambda_2 \left( \sum_{i=1}^n \sigma_i^2 - \sigma_1^2 \right) \\
& = & \frac{\lambda_1}{n} \left( \sum_{i=1}^n x_i \right)^2 + \lambda_2 \left( \sum_{i=1}^n \sigma_i^2 ||\bldv_i||_2^2 \right)
- \lambda_2 \sigma_1^2 \\
& = & \frac{\lambda_1}{n} \left( \sum_{i=1}^n x_i \right)^2 + \lambda_2 ||\bldx||_2^2 - \frac{\lambda_2}{n} \left( \sum_{i=1}^n x_i \right)^2 \; , 
\end{eqnarray*}
as claimed. 
\end{proof}

The following theorem summarizes the main result in this section. 
\begin{theorem}
Let $\code$, $\cH$, $\cH_s$ and $\cL$ be defined as above. 
Then the minimum pseudodistance with $q$-ary PSK modulation over the AWGN channel is bounded from below by 
\[
d^2_{\textrm{min},q} \ge \left( 1 - \cos ( 2 \pi / q) \right)^2 \cdot 
n \cdot \frac{2c - \lambda_2}{\lambda_1 - \lambda_2} \; . 
\] 
\label{thrm:q-ary-main}
\end{theorem}

\begin{proof}
The combination of the results in Lemmas~\ref{lemma:ineq-2-q} and~\ref{lemma:ineq-q} immediately leads to 
\[
\frac{\lambda_1 - \lambda_2}{n} \left( \sum_{i \in \cI} x_i \right)^2 + \lambda_2 || \bldx ||_2^2 
\ge 2c \cdot || \bldx ||_2^2 \; , 
\]
and by rearrangement of the coefficients we obtain 
\begin{equation}
\frac{\left( \sum_{i \in \cI} x_i \right)^2}{|| \bldx ||_2^2} \ge n \cdot \frac{2c - \lambda_2}{\lambda_1 - \lambda_2} \; . 
\label{eq:ratio-lbound-q}
\end{equation}
By re-writing~(\ref{eq:w-q}), we get 
\begin{eqnarray}
d^2_{\textrm{min},q} & \ge & \left( 1 - \cos ( 2 \pi / q) \right)^2 
\cdot \frac{\left( \sum_{i \in \cI} x_i \right)^2}{|| \bldx ||_2^2} \nonumber \\
& \ge & \left( 1 - \cos ( 2 \pi / q) \right)^2 \cdot n \frac{2c - \lambda_2}{\lambda_1 - \lambda_2} \; , 
\label{eq:main-bound}
\end{eqnarray}
where the last transition is due to~(\ref{eq:ratio-lbound-q}). 
\end{proof}

\begin{example}
Consider $\rrr = \{ 0, 1 \}$ with binary signaling over AWGN. In that case, $q =2$, and so
(\ref{eq:main-bound}) can be re-written as
\[
d^2_{\textrm{min},2} 
\ge 4 n \cdot \frac{2c - \lambda_2}{\lambda_1 - \lambda_2} \; , 
\]
which coincides with the corresponding bound in~\cite{KV-lower-bounds}, since in this case $d^2_{\textrm{eff},2}(\bldp)/4 = w_{\textrm{eff}}(\bldp)$, the \emph{effective Hamming weight} of the pseudocodeword $\bldp$. 
\end{example}
 
\begin{example}
Take $\rrr = \zz_3$ with ternary PSK over AWGN, as in Example~\ref{ex:3-1}. 
In this case, we can combine~(\ref{eq:w-3-awgn}) with~(\ref{eq:ratio-lbound-q}), thus obtaining 
\[
d^2_{\textrm{min},3} \ge 3 \cdot \frac{\left( \sum_{i \in \cI} x_i \right)^2}
{\sum_{i \in \cI} x_i^2} \ge 3 n \cdot \frac{2c - \lambda_2}{\lambda_1 - \lambda_2} \; . 
\]
Note that this bound is better then the respective bound which follows directly from~(\ref{eq:main-bound}).
\end{example}

\begin{example}
Take $\rrr = \zz_4$ with QPSK over AWGN, as in Example~\ref{ex:3-1}. 
In this case, we can combine~(\ref{eq:w-4-awgn}) with~(\ref{eq:ratio-lbound-q}), thus obtaining 
\[
d^2_{\textrm{min},4} \ge \frac{\left( \sum_{i \in \cI} x_i \right)^2}
{\sum_{i \in \cI} x_i^2} \ge n \cdot \frac{2c - \lambda_2}{\lambda_1 - \lambda_2} \; . 
\]
\end{example}

\section{Linear-Programming Bound}
\label{sec:lower-bound-2}

In this section, we present the linear-programming lower bound
on the minimum pseudodistance, similar to its counterpart 
in~\cite[Section 4]{KV-lower-bounds}.  
Throughout this section, 
let $\code$ be a linear $[n,k]$ code over $\rrr$ with an $m \times n$ parity-check matrix $\cH$,
and assume that $\cH_{j,i}$ is not a zero-divisor in $\rrr$ for any $j \in \cJ$, $i \in \cI_j$. 
Let 
\[
\cP_0 = \left( f_i(k) \right)_{i \in \cI ; k \in \{ 1,2,\cdots, q-1 \} }  
\]
be the normalized pseudocodeword matrix of a graph-cover pseudocodeword $\bldp$ corresponding to some $M$-cover of 
the Tanner graph of $\cH$. 
Denote
\[
  \forall i \in \cI \; : \; x_i =  \sum_{k=1}^{q-1} f_i(k)      \; . 
\]

It follows from Corollary~\ref{crlr:simple-inequality}, that 
\begin{equation}
\sum_{i \in \cI_j \backslash \{ \ell \} } x_i \ge x_\ell 
\label{eq:row-ineq}
\end{equation}
for all $j \in \cJ$, $\ell \in \cI_j$. 
The inequalities~(\ref{eq:row-ineq}) (for all $j \in \cJ$, $\ell \in \cI_j$)
can be expressed as 
\begin{equation}
\cK \bldx^T \ge \zeros \; ,
\end{equation} 
for some $\cK$. 

Let the entries of a vector $\bldy \in \rr^{(\cI^2)}$ be indexed by $(i,i') \in \cI^2$. 
For $i \in \cI$ we denote by $\bldy_{(i,:)}$ and $\bldy_{(:,i)}$ the sub-vectors of length $n$ 
of $\bldy$ consisting of all entries $y_{(i,i')}$ for all $i' \in \cI$ and of all entries $y_{(i',i)}$
for all $i' \in \cI$, respectively. 

The following theorem is the main result of this section. It is a generalization 
of Theorem 15 in~\cite{KV-lower-bounds}. 

\begin{theorem}
For $q$-ary PSK modulation over AWGN, the minimum pseudodistance $d_{\textrm{min},q}$ is bounded from below by
\[
d^2_{\textrm{min},q} \ge \left( 1 - \cos ( 2 \pi / q) \right)^2 \cdot \frac{1}{\max_{\bldyy \in \cK_1} \{ f'(\bldy) \} } \; , 
\]
where 
\[
f'(\bldy) = \sum_{i \in \cI} y_{(i,i)} \; , 
\]
and 
\[
\cK_1 = \left\{ \bldy \in \rr^{(\cI^2)} 
\left| \begin{array}{l}
\bldy \ge \zeros, \; \bldy \cdot \ones^T = 1 \; , \\
\cK \bldy^T_{(i,:)} \ge \zeros^T \mbox{ for all }i \in \cI \; , \\
\cK \bldy^T_{(:,i)} \ge \zeros^T \mbox{ for all }i \in \cI \; \\
\end{array} \right. 
\right\} \; . 
\]
\end{theorem}

\vspace{2ex}
{\it Sketch of the proof:}
We start with the expression in~(\ref{eq:w-q}). 
The expression 
\[
\frac{ \left( \sum_{i \in \cI} x_i \right)^2 } {\sum_{i \in \cI} x_i^2 } 
\]
can be bounded from below using the same techniques as in the proof of 
Theorem 15 in~\cite{KV-lower-bounds}, with respect to $\bldx$ defined 
as above. We omit the details.


\section*{Acknowledgements}

The authors would like to thank E. Byrne, M. Greferath and D. Sridhara for 
helpful discussions. 
This work was supported in part by the Claude Shannon Institute for Discrete Mathematics,
Coding and Cryptography (Science Foundation Ireland Grant 06/MI/006).


\end{document}